# Quantum Phase Operator and Phase States

Xin Ma

*CVS Health, Richardson, Texas 75081, USA*

William Rhodes

*Department of Chemistry, Florida State University, Tallahassee, Florida 32306, USA*

A simple solution is presented to the long-standing Dirac's phase operator problem for the quantum harmonic oscillator. A Hermitian quantum phase operator is formulated that mirrors the classical phase variable with proper time dependence and satisfies trigonometric identities. The eigenstates of the phase operator are solved in terms of Gegenbauer ultraspherical polynomials in the number state representation.

## I.  INTRODUCTION

Let $\hat{H}$ be the Hamiltonian of a quantum harmonic oscillator,

$$\hat{H} = \hat{q}^2 + \hat{p}^2 = \hat{N} + \frac{1}{2}, \qquad (\hbar\omega = 1), \tag{1.1}$$

where $\hat{q}$ and $\hat{p}$ are dimensionless position and momentum operators, $\hat{N}$ is the number operator with eigenstates $\{|n\rangle, n = 0, 1, ...\}$,

$$\hat{N} = \hat{a}^{\dagger}\hat{a}, \qquad \hat{N}|n\rangle = n|n\rangle, \tag{1.2}$$

$\hat{a}$ and $\hat{a}^{\dagger}$ are annihilation and creation operators,

$$\hat{a} = \hat{q} + i\hat{p}, \qquad \hat{a}^{\dagger} = \hat{q} - i\hat{p}, \tag{1.3}$$

with relations

$$\hat{a}|n\rangle = \sqrt{n}|n-1\rangle, \tag{1.4}$$

$$\hat{a}^{\dagger}|n\rangle = \sqrt{n+1}|n+1\rangle, \tag{1.5}$$

$$[\hat{a}, \hat{a}^{\dagger}] = 1, \qquad [\hat{a}, \hat{N}] = \hat{a}, \qquad [\hat{a}^{\dagger}, \hat{N}] = -\hat{a}^{\dagger}. \tag{1.6}$$

The missing phase operator for the quantum harmonic oscillator has been an enigma for ages. Dirac [1] in 1927 attempted to define a phase operator $\hat{\phi}$ by assuming a polar decomposition of the annihilation operator,

$$\hat{a} = e^{i\hat{\phi}} \hat{N}^{1/2}, \tag{1.7}$$

with the postulated commutation relation

$$[\hat{\phi}, \hat{N}] = -i. \tag{1.8}$$

Unfortunately, Eq. (1.7) does not define a Hermitian phase operator $\hat{\phi}$ because $e^{i\hat{\phi}}$ is not unitary and Eq. (1.8) leaves $\langle n | \hat{\phi} | n \rangle$ undefined. Louisell [2] in 1963 postulated the commutation relations

$$[\cos\hat{\phi}, \hat{N}] = i\sin\hat{\phi}, \qquad [\sin\hat{\phi}, \hat{N}] = -i\cos\hat{\phi} \tag{1.9}$$

in place of Eq. (1.8). However, Eq. (1.9) apparently does not define a Hermitian phase operator $\hat{\phi}$ either. Susskind and Glogower (SG) [3] in 1964 developed the "cosine" and "sine" operators $\hat{C}$ and $\hat{S}$. Based on the decomposition of $\hat{a}$,

$$\hat{a} = (\hat{N} + 1)^{1/2} \hat{E}, \tag{1.10}$$

where $\hat{E}$ is a "one-sided unitary" shift operator

$$\hat{E} = \sum_{n=0}^{\infty} |n\rangle\langle n+1|, \qquad \hat{E}^\dagger = \sum_{n=0}^{\infty} |n+1\rangle\langle n|, \tag{1.11}$$

with

$$\hat{E}\hat{E}^\dagger = \hat{I}, \qquad \hat{E}^\dagger \hat{E} = \hat{I} - |0\rangle\langle 0|, \tag{1.12}$$

SG defined operators $\hat{C}$ and $\hat{S}$ as

$$\hat{C} = \frac{1}{2}(\hat{E} + \hat{E}^\dagger) = \frac{1}{2}[(\hat{N}+1)^{-1/2}\hat{a} + \hat{a}^\dagger(\hat{N}+1)^{-1/2}], \tag{1.13}$$

$$\hat{S} = \frac{1}{2i}(\hat{E} - \hat{E}^\dagger) = \frac{1}{2i}[(\hat{N}+1)^{-1/2}\hat{a} - \hat{a}^\dagger(\hat{N}+1)^{-1/2}], \tag{1.14}$$

with the commutation relations equivalent to Eq. (1.9),

$$[\hat{C}, \hat{N}] = i\hat{S}, \qquad [\hat{S}, \hat{N}] = -i\hat{C}. \tag{1.15}$$

Although $\hat{C}$ and $\hat{S}$ are Hermitian, they do not commute and do not satisfy the trigonometric identity,

$$[\hat{C},\hat{S}] = \frac{1}{2i}|0\rangle\langle 0|, \tag{1.16}$$

$$\hat{C}^2 + \hat{S}^2 = \hat{I} - \frac{1}{2}|0\rangle\langle 0|. \tag{1.17}$$

The eigenvalue spectra of $\hat{C}$ and $\hat{S}$ were found by Carruthers and Nieto [4] to be continuous on $[-1,1]$,

$$\hat{C}|\cos\varphi_c\rangle = \cos\varphi_c|\cos\varphi_c\rangle, \qquad \varphi_c \in (0,\pi), \tag{1.18}$$

$$\hat{S}|\sin\varphi_s\rangle = \sin\varphi_s|\sin\varphi_s\rangle, \qquad \varphi_s \in (-\pi/2, \pi/2). \tag{1.19}$$

Pegg and Barnett (PB) [5] in 1989 proposed to define a phase operator $\hat{\phi}_s$ in an $(s+1)$-dimensional subspace $\Psi_s = \{|n\rangle, n = 0, 1, ..., s\}$ by the following polar decomposition,

$$\hat{a}_s = e^{i\hat{\phi}_s}\hat{N}_s^{1/2}, \tag{1.20}$$

where $e^{i\hat{\phi}_s}$ is a cycling operator unitary in $\Psi_s$,

$$e^{i\hat{\phi}_s} = |0\rangle\langle 1| + |1\rangle\langle 2| + \cdots + |s-1\rangle\langle s| + |s\rangle\langle 0|. \tag{1.21}$$

The PB's theory claims that one can use $\hat{\phi}_s$ to calculate physical results such as expectation values in $\Psi_s$ and take the limit $s \to \infty$ after the calculation is done. The last term $|s\rangle\langle 0|$ in Eq. (1.21) is "artificial" in order to connect the two ends of the finite spectrum of $\hat{N}_s$ to form a "loop" so that $e^{i\hat{\phi}_s}$ is unitary,

$$e^{i\hat{\phi}_s}|0\rangle = |s\rangle. \tag{1.22}$$

Eq. (1.22) not only lacks justification in physics but also creates a problem for the mathematics of PB's limiting process [6]. It is quite obvious that calculation results based on the artificial "circular" spectrum of $\hat{N}_s$ in $\Psi_s$ may not converge as $s \to \infty$. For example,

$$\lim_{s \to \infty}\langle 0|(e^{-i\hat{\phi}_s}\hat{N}_s e^{i\hat{\phi}_s})|0\rangle = \lim_{s \to \infty}\langle s|\hat{N}_s|s\rangle = \lim_{s \to \infty} s = \infty. \tag{1.23}$$

As a matter of fact, the PB formalism is equivalent to what Louisell and Gordon proposed in 1961 [7-8].

In this paper, we define a Hermitian phase operator that properly mirrors the classical phase and is free of the problems encountered by various approaches based on polar decomposition of $\hat{a}$ due to the one-sidedness of the number operator spectrum [9]. The definition of the phase operator and the solutions of its eigenstates are given in Sec. II. The properties of the phase operator are analyzed in Sec. III.

## II. FORMULATION OF PHASE OPERATOR

A proper phase operator $\hat{\phi}$ for the quantum harmonic oscillator should resemble the classical phase variable and is expected to have the following desired properties.

1) $\hat{\phi}$ is Hermitian in the infinite-dimensional Hilbert space $\mathbf{H}$,

$$\hat{\phi}^{\dagger} = \hat{\phi}, \quad \hat{\phi}|\psi\rangle \in \mathbf{H}, \quad \forall |\psi\rangle \in \mathbf{H} = \{|n\rangle, n = 0, 1, ...\}. \tag{2.1}$$

2) $\hat{\phi}$ satisfies the trigonometric identity,

$$\cos^2 \hat{\phi} + \sin^2 \hat{\phi} = \hat{I}. \tag{2.2}$$

3) $\hat{\phi}$ has proper time dependence. Equation $\hat{\phi}(t) = \hat{\phi}(0) - \omega t$ is not expected to hold true at all times in the Heisenberg picture (otherwise $\hat{\phi}(t)$ would effectively be a time operator). However, it should hold true at least at the half period time $t = \pi / \omega$ to reflect the point of inversion when $\hat{q}(\pi/\omega) = -\hat{q}(0)$ and $\hat{p}(\pi/\omega) = -\hat{p}(0)$,

$$\hat{\phi}(\frac{\pi}{\omega}) = \hat{\phi}(0) - \pi. \tag{2.3}$$

In other words, an eigenstate $|\varphi\rangle$ of $\hat{\phi}$ evolves into the eigenstate $|\varphi - \pi\rangle$ at $t = \pi/\omega$.

### A. Definition of phase operator

We define our phase operator $\hat{\phi}$ such that

$$\cos^2 \hat{\phi} = (\hat{N} + \frac{1}{2})^{-1/2} \hat{q}^2 (\hat{N} + \frac{1}{2})^{-1/2}, \tag{2.4}$$

$$\sin^2 \hat{\phi} = (\hat{N} + \frac{1}{2})^{-1/2} \hat{p}^2 (\hat{N} + \frac{1}{2})^{-1/2}, \tag{2.5}$$

where

$$\cos \hat{\phi} = \sum_{n=0}^{\infty} (-1)^n \frac{\hat{\phi}^{2n}}{(2n)!}, \quad \sin \hat{\phi} = \sum_{n=0}^{\infty} (-1)^n \frac{\hat{\phi}^{2n+1}}{(2n+1)!}. \tag{2.6}$$

Our definitions of $\cos^2 \hat{\phi}$ and $\sin^2 \hat{\phi}$ are completely consistent with $\cos^2 \phi$ and $\sin^2 \phi$ of the classical phase variable $\phi$, representing the percentages of the total energy distributed to position and momentum respectively at any time. Clearly, our phase operator $\hat{\phi}$ is Hermitian and satisfies the trigonometric identity (2.2).

From Eq. (1.3) and (2.4) - (2.5) we find

$$\cos 2\hat{\phi} = \cos^2 \hat{\phi} - \sin^2 \hat{\phi} = \frac{1}{2}(\hat{N}+\frac{1}{2})^{-1/2}(\hat{a}^2 + \hat{a}^{\dagger 2})(\hat{N}+\frac{1}{2})^{-1/2}. \tag{2.7}$$

Applying Eq. (1.4) - (1.5) we have

$$(\hat{N}+\frac{1}{2})^{-1/2}\hat{a}^2(\hat{N}+\frac{1}{2})^{-1/2}|n\rangle = \sqrt{f(n-1)}\,|n-2\rangle, \tag{2.8}$$

$$(\hat{N}+\frac{1}{2})^{-1/2}\hat{a}^{\dagger 2}(\hat{N}+\frac{1}{2})^{-1/2}|n\rangle = \sqrt{f(n+1)}\,|n+2\rangle, \tag{2.9}$$

where

$$f(n) = \begin{cases} \dfrac{n(n+1)}{(n-1/2)(n+3/2)}, & n = 1, 2, ..., \\ 0, & n \leq 0, \end{cases} \tag{2.10}$$

which decreases asymptotically to 1 for $n \geq 1$,

$$\frac{8}{5} \geq f(n) > 1, \qquad \lim_{n \to \infty} f(n) = 1. \tag{2.11}$$

From Eq. (1.6) it is straightforward to find the following commutation relations

$$[\cos 2\hat{\phi}, \hat{N}] = (\hat{N}+\frac{1}{2})^{-1/2}(\hat{a}^2 - \hat{a}^{\dagger 2})(\hat{N}+\frac{1}{2})^{-1/2}, \tag{2.12}$$

$$[\cos^2 \hat{\phi}, \hat{N}] = \frac{1}{2}[\cos 2\hat{\phi}, \hat{N}], \tag{2.13}$$

$$[\sin^2 \hat{\phi}, \hat{N}] = -\frac{1}{2}[\cos 2\hat{\phi}, \hat{N}]. \tag{2.14}$$

Note that simply taking the inverse cosine of Eq. (2.7) does not give a complete definition of $\hat{\phi}$. We will give explicit definition of $\hat{\phi}$ in the next subsection with its time dependence taken into consideration.

## B. Eigenstates and eigenvalues

Let $|\lambda\rangle$ be an eigenstate of $\cos 2\hat{\phi}$,

$$\cos 2\hat{\phi}|\lambda\rangle = \lambda|\lambda\rangle. \qquad (2.15)$$

Expanding $|\lambda\rangle$ in the number states $\{|n\rangle\}$ and applying Eq. (2.7) - (2.9), we have

$$\cos 2\hat{\phi}|\lambda\rangle = \frac{1}{2}\sum_{n=0}^{\infty}\left(\sqrt{f(n-1)}|n-2\rangle + \sqrt{f(n+1)}|n+2\rangle\right)\langle n|\lambda\rangle = \lambda\sum_{n=0}^{\infty}|n\rangle\langle n|\lambda\rangle, \qquad (2.16)$$

Comparing the coefficient of $|n\rangle$ in Eq. (2.16), we find the following recurrence relation for $\langle n|\lambda\rangle$,

$$\sqrt{f(n+1)}\langle n+2|\lambda\rangle = 2\lambda\langle n|\lambda\rangle - \sqrt{f(n-1)}\langle n-2|\lambda\rangle, \quad n=0,1,\ldots. \qquad (2.17)$$

Now let us decompose the Hilbert space $\mathbf{H}$ into a direct sum of two orthogonal subspaces,

$$\mathbf{H} = \mathbf{H}_e \oplus \mathbf{H}_o, \qquad (2.18)$$

where $\mathbf{H}_e$ and $\mathbf{H}_o$ are spanned by even and odd number states respectively,

$$\mathbf{H}_e = \{|0\rangle,|2\rangle,\ldots\}, \qquad \mathbf{H}_o = \{|1\rangle,|3\rangle,\ldots\}. \qquad (2.19)$$

We can then write

$$|\lambda\rangle = |\lambda,e\rangle + |\lambda,o\rangle, \quad |\lambda,e\rangle \in \mathbf{H}_e, \quad |\lambda,o\rangle \in \mathbf{H}_o. \qquad (2.20)$$

Since $\mathbf{H}_e$ and $\mathbf{H}_o$ are closures of $\cos 2\hat{\phi}$ from Eq. (2.7) - (2.9), namely,

$$\cos 2\hat{\phi}|\lambda,e\rangle \in \mathbf{H}_e, \qquad \cos 2\hat{\phi}|\lambda,o\rangle \in \mathbf{H}_o, \qquad (2.21)$$

both $|\lambda,e\rangle$ and $|\lambda,o\rangle$ are eigenstates of $\cos 2\hat{\phi}$ with eigenvalue $\lambda$. We therefore have

$$\cos 2\hat{\phi}[|\lambda,e\rangle,|\lambda,o\rangle] = [|\lambda,e\rangle,|\lambda,o\rangle]\begin{bmatrix}\lambda & 0 \\ 0 & \lambda\end{bmatrix}. \qquad (2.22)$$

For even and odd number states respectively, the recurrence relation (2.17) becomes

$$\sqrt{f(2n+1)}\langle 2(n+1)|\lambda,e\rangle = 2\lambda\langle 2n|\lambda,e\rangle - \sqrt{f(2n-1)}\langle 2(n-1)|\lambda,e\rangle, \qquad (2.23)$$

and

$$\sqrt{f(2n+2)}\langle 2(n+1)+1|\lambda,o\rangle = 2\lambda\langle 2n+1|\lambda,o\rangle - \sqrt{f(2n)}\langle 2(n-1)+1|\lambda,o\rangle. \qquad (2.24)$$

The solutions to the recurrence relations (2.23) - (2.24) are straightforward,

$$\langle 2n | \lambda, e \rangle = N_n^{-1}(1/4)(1-\lambda^2)^{-1/8} C_n^{(1/4)}(\lambda), \qquad n = 0, 1, ..., \tag{2.25}$$

$$\langle 2n+1 | \lambda, o \rangle = N_n^{-1}(3/4)(1-\lambda^2)^{1/8} C_n^{(3/4)}(\lambda), \qquad n = 0, 1, ..., \tag{2.26}$$

with

$$\langle \lambda, e | \lambda', e \rangle = \langle \lambda, o | \lambda', o \rangle = \delta(\lambda - \lambda'), \qquad \langle \lambda, e | \lambda', o \rangle = 0, \tag{2.27}$$

where $C_n^{(\alpha)}(\lambda)$ are Gegenbauer ultraspherical polynomials orthogonal on $[-1,1]$ with respect to the weight function $(1-\lambda^2)^{\alpha-1/2}$ for fixed $\alpha > -1/2$,

$$\int_{-1}^{1}(1-\lambda^2)^{\alpha-1/2} C_m^{(\alpha)}(\lambda) C_n^{(\alpha)}(\lambda) d\lambda = N_n^2(\alpha)\delta_{mn}, \tag{2.28}$$

and

$$N_n^2(\alpha) = \frac{2^{1-2\alpha}\pi\Gamma(n+2\alpha)}{(n+\alpha)n!\Gamma^2(\alpha)}. \tag{2.29}$$

Since $[\cos 2\hat{\phi}, \hat{\phi}] = 0$ by definition (2.6), $\cos 2\hat{\phi}$ and $\hat{\phi}$ have a common set of eigenstates. Let $\{|\lambda, +\rangle, |\lambda, -\rangle, \lambda \in (-1,1)\}$ be the common set of eigenstates of both $\cos 2\hat{\phi}$ and $\hat{\phi}$ defined by the following unitary transformation,

$$[|\lambda, +\rangle, |\lambda, -\rangle] = [|\lambda, e\rangle, |\lambda, o\rangle] \frac{1}{\sqrt{2}}\begin{bmatrix} 1 & 1 \\ 1 & -1 \end{bmatrix}, \qquad \lambda \in (-1,1), \tag{2.30}$$

with

$$\langle \lambda, + | \lambda', + \rangle = \langle \lambda, - | \lambda', - \rangle = \delta(\lambda - \lambda'), \qquad \langle \lambda, + | \lambda', - \rangle = 0. \tag{2.31}$$

We define our phase operator $\hat{\phi}$ such that

$$\hat{\phi}[|\lambda, +\rangle, |\lambda, -\rangle] = [|\lambda, +\rangle, |\lambda, -\rangle]\begin{bmatrix} \varphi & 0 \\ 0 & \varphi - \pi \end{bmatrix}, \qquad \varphi = \frac{1}{2}\cos^{-1}\lambda, \quad \forall \lambda \in (-1,1). \tag{2.32}$$

Eigenstates $|\lambda, +\rangle$ and $|\lambda, -\rangle$ will be referred to as phase states because they evolve into each other periodically and are out of phase by $\pi$ as we will show in Sec. III.

From Eq. (2.32) any analytic functions of $\hat{\phi}$ can be easily expressed in the $[|\lambda, +\rangle, |\lambda, -\rangle]$ basis. For example, $\tan\hat{\phi}$ corresponding to the classical expression $\tan\phi = p/q$ is given by

$$\tan\hat{\phi}\left[|\lambda,+\rangle,|\lambda,-\rangle\right]=\left[|\lambda,+\rangle,|\lambda,-\rangle\right]\begin{bmatrix}\tan\varphi & 0\\ 0 & \tan\varphi\end{bmatrix},\quad \varphi=\frac{1}{2}\cos^{-1}\lambda,\quad \forall\lambda\in(-1,1). \tag{2.33}$$

For convenience, we can use notation $|\varphi\rangle$ to collectively denote phase states $|\lambda,+\rangle$ and $|\lambda,-\rangle$,

$$|\varphi\rangle=\begin{cases}\sqrt{2}(1-\lambda^2)^{1/4}|\lambda,+\rangle, & \varphi\in(0,\pi/2),\\ \sqrt{2}(1-\lambda^2)^{1/4}|\lambda,-\rangle, & \varphi\in(\pi,3\pi/2),\end{cases}\quad \cos 2\varphi\equiv\lambda\in(-1,1), \tag{2.34}$$

with $|\varphi\rangle$ normalized to a δ-function on the angular spectrum,

$$\langle\varphi|\varphi'\rangle=\delta(\varphi-\varphi'). \tag{2.35}$$

The identity operator can be resolved as

$$\hat{I}=\int_{-1}^{1}d\lambda\,|\lambda,+\rangle\langle\lambda,+|+\int_{-1}^{1}d\lambda\,|\lambda,-\rangle\langle\lambda,-|=\int_{0}^{\frac{\pi}{2}}d\varphi\,|\varphi\rangle\langle\varphi|+\int_{\pi}^{\frac{3\pi}{2}}d\varphi\,|\varphi\rangle\langle\varphi|. \tag{2.36}$$

The spectrum representation of $\hat{\phi}$ therefore is

$$\hat{\phi}=\int_{0}^{\frac{\pi}{2}}\varphi d\varphi\,|\varphi\rangle\langle\varphi|+\int_{\pi}^{\frac{3\pi}{2}}\varphi d\varphi\,|\varphi\rangle\langle\varphi|. \tag{2.37}$$

With the explicit definition $\hat{\phi}$, our $\cos\hat{\phi}$ and $\sin\hat{\phi}$ can be viewed as a normalization of SG's $\hat{C}$ and $\hat{S}$ operators without the problems in Eq. (1.16) - (1.17).

## III. PROPERTIES OF PHASE OPERATOR

### A. Time evolution of phase operator

The time evolution of phase states $|\lambda,+\rangle$ and $|\lambda,-\rangle$ is determined by the following equations (let $\omega=1$ for notational convenience),

$$e^{-it\hat{H}}|\lambda,+\rangle=\frac{1}{\sqrt{2}}\left(\sum_{n=0}^{\infty}e^{-i2nt}|2n\rangle\langle 2n|\lambda,e\rangle+e^{-it}\sum_{n=0}^{\infty}e^{-i2nt}|2n+1\rangle\langle 2n+1|\lambda,o\rangle\right), \tag{3.1}$$

$$e^{-it\hat{H}}|\lambda,-\rangle=\frac{1}{\sqrt{2}}\left(\sum_{n=0}^{\infty}e^{-i2nt}|2n\rangle\langle 2n|\lambda,e\rangle-e^{-it}\sum_{n=0}^{\infty}e^{-i2nt}|2n+1\rangle\langle 2n+1|\lambda,o\rangle\right). \tag{3.2}$$

At times $t=\pi/2,\ \pi,\ 3\pi/2$, we have

$$e^{-i\frac{\pi}{2}\hat{H}}\left[|\lambda,+\rangle,|\lambda,-\rangle\right]=\left[|-\lambda,+\rangle,|-\lambda,-\rangle\right]\frac{1}{2}\begin{bmatrix}1-i & 1+i \\ 1+i & 1-i\end{bmatrix}, \quad (3.3)$$

$$e^{-i\pi\hat{H}}\left[|\lambda,+\rangle,|\lambda,-\rangle\right]=\left[|\lambda,+\rangle,|\lambda,-\rangle\right]\begin{bmatrix}0 & 1 \\ 1 & 0\end{bmatrix}, \quad (3.4)$$

$$e^{-i\frac{3\pi}{2}\hat{H}}\left[|\lambda,+\rangle,|\lambda,-\rangle\right]=\left[|-\lambda,+\rangle,|-\lambda,-\rangle\right]\frac{1}{2}\begin{bmatrix}1+i & 1-i \\ 1-i & 1+i\end{bmatrix}. \quad (3.5)$$

Eq. (3.4) shows that $|\lambda,+\rangle$ and $|\lambda,-\rangle$ evolves into each other at time $t=\pi$.

From Eq. (3.3) - (3.5), $\hat{\phi}(t)=e^{it\hat{H}}\hat{\phi}(0)e^{-it\hat{H}}$ at times $t=\pi/2,\ \pi,\ 3\pi/2$ can be expressed as

$$\hat{\phi}(\frac{\pi}{2})\left[|\lambda,+\rangle,|\lambda,-\rangle\right]=\left[|\lambda,+\rangle,|\lambda,-\rangle\right]\begin{bmatrix}\pi-\varphi & -i\pi/2 \\ i\pi/2 & \pi-\varphi\end{bmatrix}, \quad (3.6)$$

$$\hat{\phi}(\pi)\left[|\lambda,+\rangle,|\lambda,-\rangle\right]=\left[|\lambda,+\rangle,|\lambda,-\rangle\right]\begin{bmatrix}\varphi-\pi & 0 \\ 0 & \varphi\end{bmatrix}, \quad (3.7)$$

$$\hat{\phi}(\frac{3\pi}{2})\left[|\lambda,+\rangle,|\lambda,-\rangle\right]=\left[|\lambda,+\rangle,|\lambda,-\rangle\right]\begin{bmatrix}\pi-\varphi & i\pi/2 \\ -i\pi/2 & \pi-\varphi\end{bmatrix}. \quad (3.8)$$

Note that Eq. (3.7) is equivalent to Eq. (2.3) confirming $\hat{\phi}(t)$ is a phase shift operator by $\pi$ at time $t=\pi$.

Similarly, for $\cos\hat{\phi}(t)$, $\cos^2\hat{\phi}(t)$, and $\cos 2\hat{\phi}(t)$ at times $t=\pi/2,\ \pi,\ 3\pi/2$, we have

$$\cos\hat{\phi}(\frac{\pi}{2})\left[|\lambda,+\rangle,|\lambda,-\rangle\right]=\left[|\lambda,+\rangle,|\lambda,-\rangle\right]\begin{bmatrix}0 & i\sin\varphi \\ -i\sin\varphi & 0\end{bmatrix}, \quad (3.9)$$

$$\cos\hat{\phi}(\pi)=-\cos\hat{\phi}(0), \quad (3.10)$$

$$\cos\hat{\phi}(\frac{3\pi}{2})\left[|\lambda,+\rangle,|\lambda,-\rangle\right]=\left[|\lambda,+\rangle,|\lambda,-\rangle\right]\begin{bmatrix}0 & -i\sin\varphi \\ i\sin\varphi & 0\end{bmatrix}, \quad (3.11)$$

$$\cos^2\hat{\phi}(\frac{\pi}{2})=\sin^2\hat{\phi}(0),\quad \cos^2\hat{\phi}(\pi)=\cos^2\hat{\phi}(0),\quad \cos^2\hat{\phi}(\frac{3\pi}{2})=\sin^2\hat{\phi}(0), \quad (3.12)$$

$$\cos 2\hat{\phi}(\frac{\pi}{2})=-\cos 2\hat{\phi}(0),\quad \cos 2\hat{\phi}(\pi)=\cos 2\hat{\phi}(0),\quad \cos 2\hat{\phi}(\frac{3\pi}{2})=-\cos 2\hat{\phi}(0). \quad (3.13)$$

Similar expressions for $\sin\hat{\phi}(t)$, $\sin^2\hat{\phi}(t)$, and $\sin 2\hat{\phi}(t)$ can be easily found by swapping "cos" and "sin" in Eq. (3.9) - (3.13). Although $\hat{\phi}(t), \cos\hat{\phi}(t), \sin\hat{\phi}(t)$ are not diagonal at $t=\pi/2$ and $t=3\pi/2$, $\cos^2\hat{\phi}(t), \sin^2\hat{\phi}(t)$ and thus $\cos 2\hat{\phi}(t)$, $\sin 2\hat{\phi}(t)$ are. Therefore, in phase states $|\varphi\rangle$, the alternating eigenvalue of $\cos 2\hat{\phi}$ between $\lambda$ and $-\lambda$ every quarter period implies that phase information is at the granularity of quarter period (as opposed to being infinitely fine with a classical harmonic oscillator).

## B. Phase distributions in number states

It is easy to show that in any number state $|n\rangle$ $\cos\hat{\phi}$ and $\sin\hat{\phi}$ have exactly the same moments,

$$m_{2k+1}(n) = \langle n|\cos^{2k+1}\hat{\phi}|n\rangle = \langle n|\sin^{2k+1}\hat{\phi}|n\rangle = 0, \qquad k=0,1,..., \quad (3.14)$$

$$m_{2k}(n) = \langle n|\cos^{2k}\hat{\phi}|n\rangle = \langle n|\sin^{2k}\hat{\phi}|n\rangle = \frac{1}{2^k}\sum_{m=0}^{[k/2]} C_k^{2m}\langle n|\cos^{2m}2\hat{\phi}|n\rangle, \qquad k=1,2,..., \quad (3.15)$$

where $C_k^{2m}$ are binomial coefficients.

For example,

$$m_2(n) = \frac{1}{2}, \tag{3.16}$$

$$m_4(n) = \frac{1}{4} + \frac{1}{16}[f(n+1) + f(n-1)], \tag{3.17}$$

$$m_6(n) = \frac{1}{8} + \frac{3}{32}[f(n+1) + f(n-1)]. \tag{3.18}$$

From Eq. (2.10) - (2.11) we conclude that

$$m_{2k}(n) \leq u_{2k}, \ n=0,1, \qquad m_{2k}(n) \geq u_{2k}, \ n=2,3,..., \tag{3.19}$$

where the equal sign applies when $k=1$ and $u_{2k}$ are moments of uniform distribution,

$$u_{2k} = <\cos^{2k}\varphi> = <\sin^{2k}\varphi> = \frac{(2k-1)!!}{(2k)!!}, \quad k=1,2,.... \tag{3.20}$$

With $u_4 = 3/8$ and $u_6 = 5/16$, we have

$$m_4(0) = \frac{7}{20} < u_4, \quad m_4(1) = \frac{9}{28} < u_4, \quad m_4(2) = \frac{5}{12} > u_4, \tag{3.21}$$

$$m_6(0) = \frac{11}{40} < u_6, \quad m_6(1) = \frac{13}{56} < u_6, \quad m_6(2) = \frac{3}{8} > u_6. \tag{3.22}$$

The reason we have $m_{2k} \leq u_{2k}$ for the two bottom number states $\{|0\rangle, |1\rangle\}$ is that $f(n-1)$ in Eq. (2.8) vanishes giving no contribution to the sum in Eq. (3.15) due to the one-sided boundedness of the number operator $\hat{N}$. This can be easily seen in Eq. (3.17) - (3.18). As Carruthers and Nieto [4] showed in 1968, quantum phase in general is not of uniform distribution in number states $|n\rangle$ contrary to common belief.

In the limit of $n \to \infty$, we have

$$\lim_{n \to \infty} m_{2k}(n) = \frac{1}{2^k} \sum_{m=0}^{[k/2]} \frac{1}{2^{2m}} C_k^{2m} C_{2m}^m = \frac{(2k-1)!!}{(2k)!!} = u_{2k}, \quad k = 1, 2, \ldots, \tag{3.23}$$

confirming that number states $|n\rangle$ approach uniform phase distribution as $n \to \infty$.

## C. Phase in coherent states

In a coherent state $|\alpha\rangle$, where $\alpha = |\alpha|e^{i\phi}$, the expectation value of $\cos 2\hat{\phi}$ is

$$\langle \alpha | \cos 2\hat{\phi} | \alpha \rangle = e^{-|\alpha|^2} \cos 2\phi \sum_{n=0}^{\infty} \frac{|\alpha|^{2(n+1)}}{n! \sqrt{(n+5/2)(n+1/2)}}, \tag{3.24}$$

From Eq. (2.12) and Eq. (3.24) it immediately follows that

$$\langle \alpha | -i[\cos 2\hat{\phi}, \hat{H}] | \alpha \rangle = e^{-|\alpha|^2} 2 \sin 2\phi \sum_{n=0}^{\infty} \frac{|\alpha|^{2(n+1)}}{n! \sqrt{(n+5/2)(n+1/2)}}. \tag{3.25}$$

In highly excited coherent states which is the classical limit of quantum states, we have

$$\lim_{|\alpha| \to \infty} \langle \alpha | \cos 2\hat{\phi} | \alpha \rangle = \cos 2\phi. \tag{3.26}$$

The expectation values of the following commutators in the classical limit are exactly the same as the corresponding classical Poisson brackets (on the right hand side),

$$\lim_{|\alpha| \to \infty} \langle \alpha | -i[\cos 2\hat{\phi}, \hat{H}] | \alpha \rangle = 2 \sin 2\phi = \{\cos 2\phi, H\}, \tag{3.27}$$

$$\lim_{|\alpha| \to \infty} \langle \alpha | -i[\cos^2 \hat{\phi}, \hat{H}] | \alpha \rangle = \sin 2\phi = \{\cos^2 \phi, H\}, \tag{3.28}$$

$$\lim_{|\alpha| \to \infty} \langle \alpha | -i[\sin^2 \hat{\phi}, \hat{H}] | \alpha \rangle = -\sin 2\phi = \{\sin^2 \phi, H\}. \tag{3.29}$$

Eq. (3.26) - (3.29) confirm that our phase operator $\hat{\phi}$ has correct behavior in the classical limit.

## ACKNOWLEDGMENT

This work was supported by Contract No. DE-FG05-86ER60473 between the Division of Biomedical and Environmental Research of the Department of Energy and Florida State University.